\documentclass[prd,nofootinbib,unsortedaddress,superscriptaddress,showpacs]{revtex4-1}

\usepackage{amsfonts}
\usepackage{amsmath}
\usepackage{amssymb}
\usepackage{bbm}
\usepackage{graphics}

\newcommand{\kM}{$\kappa$-Minkowski }

\newcommand{\Lag}{\mathcal{L}}

\newcommand{\G}{\mathcal{G}}

\newcommand{\g}{\mathfrak{g}}

\newcommand{\Mtheta}{\mathcal{M}_\theta}

\newcommand{\CG}{\mathbb{C}[\G]}
\newcommand{\CM}{\mathbb{C}[\mathcal{M}]}
\newcommand{\CqG}{\mathbb{C}_q[\G]}
\newcommand{\CqM}{\mathbb{C}_q[\mathcal{M}]}

\newcommand{\id}{\mathbbm{1}}

\newcommand{\lbr}{\text{[}}
\newcommand{\rbr}{\text{]}}

\begin{document}

\title{Conserved charges and quantum-group transformations\\
 in noncommutative field theories}

\author{Giovanni AMELINO-CAMELIA}
\email{giovanni.amelino-camelia@roma1.infn.it}
\affiliation{Dipartimento di Fisica \\
 Universit\`a di Roma ``La Sapienza"\\
and Sez.~Roma1 INFN\\
P.le A. Moro 2, 00185 Roma , Italy}

\author{Giulia GUBITOSI}
\affiliation{Dipartimento di Fisica \\
 Universit\`a di Roma ``La Sapienza"\\
and Sez.~Roma1 INFN\\
P.le A. Moro 2, 00185 Roma , Italy}

\author{Flavio MERCATI}
\affiliation{Dipartimento di Fisica \\
 Universit\`a di Roma ``La Sapienza"\\
and Sez.~Roma1 INFN\\
P.le A. Moro 2, 00185 Roma , Italy}

\author{Giacomo ROSATI}
\affiliation{Dipartimento di Fisica \\
 Universit\`a di Roma ``La Sapienza"\\
and Sez.~Roma1 INFN\\
P.le A. Moro 2, 00185 Roma , Italy}

\begin{abstract}
The recently-developed techniques of Noether analysis
of the quantum-group spacetime symmetries of some noncommutative field theories
rely on the {\it ad hoc} introduction of some peculiar auxiliary transformation
parameters, which appear to have no role in the structure of the quantum group.
We here show that it is possible to set up the Noether analysis
 directly in terms of the quantum-group symmetry transformations,
  and we therefore establish more robustly the attribution of the
conserved charges to the symmetries of interest.
We also characterize the concept of ``time independence"
(as needed for conserved charges)
in a way that is robust enough to be applicable even
 to theories
with space/time noncommutativity, where it might have appeared
that any characterization of
time independence should be vulnerable to changes of ordering convention.
\end{abstract}

\maketitle

\tableofcontents

\section{Introduction and summary}\label{intro}
Over the last decade there has been a strong research effort
focused on theories formulated in noncommutative
versions of Minkowski spacetime.
Among the reasons of interest in such studies several concern the
implications of the noncommutativity of spacetime coordinates for the fate
of spacetime symmetries (see, {\it e.g.}, Ref.~\cite{dopliPLB,majrue,kpoinap,gacmaj,chaichan,wessfiore,balaIRUV}),
an issue which is not only of obvious conceptual appeal but also provides the basis
for an intriguing phenomenological program (see, {\it e.g.}, Ref.~\cite{grbgac,gacdsr,iruv1,iruv2,iruv4,dsrnature,orfeupion,repBertolami}).
The simplest and most studied possibility
for spacetime noncommutativity
are ``canonical spacetimes"~\cite{dopliPLB,chaichan,wessfiore,balaIRUV,wessNCYM,thetanoether}
(or ``$\theta$-Minkowski") with coordinates such
that\footnote{We use conventions for spacetime indices such
that $\mu , \nu \in \{ 0,1,2,3 \}$, $j , k \in \{ 1,2,3 \}$, and $x_0$ is understood
as time coordinate.}
\begin{equation}
 [ x_\mu , x_\nu ] = i \theta_{\mu\nu} ~.
\label{canonical}
\end{equation}
Even restricting one's attention to this possibility
the literature offers a multitude of alternative scenarios.
The research program initiated in Ref.~\cite{dopliPLB} intends to find suitable
restrictions on the form of $\theta_{\mu \nu}$ and to introduce appropriate
nontrivial algebraic properties~\cite{dopliFUZZYTHETA} to $\theta_{\mu \nu}$.
In recent times most of the related
literature has focused on the less ambitious
possibility of a (dimensionful) number-valued matrix $\theta_{\mu \nu}$, 
and this in turn splits
very sharply into two alternative scenarios.
The conceptually simplest scenario assumes that $\theta_{\mu \nu}$
 behaves like
a tensor under ordinary Lorentz/Poincar\'e transformations, and as a result it predicts
a breakdown of relativistic properties somewhat analogous to the case of light propagation
in anisotropic media: the tensor $\theta_{\mu \nu}$ takes of course different values
for different inertial observers and this breaks the relativistic equivalence of
inertial frames.
The other much studied possibility assumes that $\theta_{\mu \nu}$
is a constant/invariant matrix, whose matrix elements take exactly the same numerical value
in all inertial frames, and as a result (in light of the relationship between
products of coordinates and the noncommutativity matrix) is incompatible with symmetry
under classical Lorentz transformations.
It turns out that, as we shall summarize in  Sections \ref{introTHETA}
and \ref{quantumgroup}, in order
to have  the noncommutativity matrix as a relativistic invariant it is necessary
to describe the laws of transformation between inertial frames in terms of
a Hopf algebra.

As several recent studies we shall here focus on this latter
possibility, with relativistic symmetries described by a Hopf algebra.
The idea of ``deformation" (rather than breakdown)
of spacetime symmetries in a quantum spacetime, preliminarily
proposed in Refs.~\cite{gacdsr,dsrnature},
is attracting interest as a plausible scenario for
quantum-gravity research~\cite{carloDSR,leeDSRlectures}.
And theories in noncommutative spacetime with Hopf-algebra symmetries
could be a valuable laboratory for sharpening these novel concepts.
Indeed several studies
have adopted this perspective, focusing both on $\theta$-Minkowski,
of (\ref{canonical}),
and on ``$\kappa$-Minkowski" noncommutative
 spacetime~\cite{majrue,kpoinap,gacmaj,kappanoether,nopure},
with the characteristic noncommutativity
\begin{eqnarray}
[x_j,x_0]= i \, \lambda \, x_j~,~~~[x_k,x_j]=0 ~. \label{kmnoncomm}
\end{eqnarray}

We are mainly concerned here with some of the unsettled issues that need to be
clarified in order to establish whether the relevant Hopf-algebra spacetime symmetries
are strong enough to produce conserved currents and charges.
Some techniques of Noether analysis of
these novel symmetries were only developed very
recently~\cite{kappanoether,jurekNOETHER1,jurekNOETHER2,thetanoether,nopure},
but the interpretation of the currents and charges produced by these techniques
has not yet been fully clarified.
Part of the residual concerns are due to the fact that these
recent Noether analyses found necessary to introduce
some  {\it ad hoc} ``infinitesimal noncommutative
transformation parameters", which puzzlingly could not be expressed in terms
of previously known mathematical properties of the relevant Hopf algebras.
This clearly may give rise to some skepticism concerning
  the attribution of the
conserved charges to the symmetries of interest.
Our main goal here is to show that there is no need for such peculiarities:
the Noether analysis of the relevant Hopf-algebra symmetries can be performed
following exactly the same steps of the Noether analysis of classical symmetries,
of course replacing the classical-symmetry rules of transformation of fields with
the Hopf-algebra transformation rules. This is what we
accomplish in Section \ref{noether},
where we rely on a standard description of the quantum-group symmetry transformations,
which for $\theta$-Minkowski involves the ``twisted Poincar\'e" (or ``$\theta$-Poincar\'e)
 quantum group \cite{chaichan,KosinskiThetaPoincarGroup,oecklTHETA}
(here reviewed in Section \ref{quantumgroup}).

After having established more firmly the attribution of the conserved currents
and charges to the Hopf-algebra spacetime symmetries, we contemplate in Section \ref{timeindep}
the issue of how to properly introduce the concept of time-independent quantities
(like conserved charges) and in general of stationary fields in theories
with ``space/time noncommutativity", where the time coordinate
is noncommutative and its presence in an expression may appear
to depend on ordering conventions. We introduce a robust
(ordering-convention independent) characterization of
stationary fields (and conserved charges) which we show to be fully satisfactory.

 In the closing Section \ref{closing} we offer some remarks on possible applications
 of our findings and a perspective on other challenges that deserve priority
 in this research area.

\section{Auxiliary noncommutative infinitesimal
transformation parameters}\label{introTHETA}
We find convenient to first briefly summarize the description of Hopf-algebra
symmetries and
 the peculiarities of
 the ``noncommutative infinitesimal transformation parameters" used in the
 Noether analysis~\cite{thetanoether}  for $\theta$-Minkowski (which are completely analogous
to the type of parameters used for Noether analyses in $\kappa$-Minkowski
spacetime~\cite{kappanoether,jurekNOETHER1,jurekNOETHER2,thetanoether,nopure}).

It is convenient~\cite{wessNCYM} to describe fields in $\theta$-Minkowski in terms
of a basis of exponentials
\begin{equation}
\Phi(x) = \int d^4 k \,\tilde \Phi(k) e^{i k^\mu x_\mu} ~,
\label{fourierW}
\end{equation}
where the ``Fourier parameters" $k_\mu$ are commutative~\cite{wessNCYM}.
The novel properties of these fields are conveniently all encoded in the $x_\mu$ dependence
of the basis of exponentials, and all ordering issues are taken care of by specifying the basis.
In (\ref{fourierW}) we adopted (as most frequently preferred in the literature)
the ``Weyl basis" $e^{i k^\mu x_\mu}$. Clearly one may choose among
infinitely many alternative ordering conventions, such as the time-to-the-right
 ordering convention $e^{i k^j x_j} e^{i k^0 x_0}$, and
a change of ordering prescription for the basis of exponentials
clearly demands a change of ``Fourier transform" $\tilde \Phi(k)$ in order to describe
the same field. But of course the ``physical predictions" of such field theories
will be ultimately independent~\cite{thetanoether} of this choice of ordering convention.

One then introduces rules of ``classical action" of symmetry generators on basis elements:
\begin{equation}
 P_\mu \triangleright e^{i k x} = - k_\mu e^{i k x} ~,
 \label{actionP}
\end{equation}
for translation generators, and\footnote{We adopt a commonly-used notation, with
(anti)symmetrization with respect to pairs of indices denoted by (square)curly brackets.}
\begin{equation}
  M_{\mu\nu} \triangleright e^{i k x} = \tfrac{1}{2} \left[ x_{[\mu} ( P_{\nu]} \triangleright e^{i k x})+ ( P_{[\nu} \triangleright e^{i k x})  x_{\mu]}  \right]= - \tfrac{1}{2} k_{[\nu} \left[ x_{\mu]} e^{i k x} + e^{i k x} x_{\mu]}\right] ~,
\label{actionM}
\end{equation}
for Lorentz-sector generators.
Of course then commutators of these generators are the same as in
 the ordinary Poincar\'e Lie algebra:
\begin{eqnarray}
\left[ P_\mu , P_\nu \right]&=&0 \nonumber\\
\left[ P_\alpha,M_{\mu\nu}  \right]&=& i \eta_{\alpha [\mu} P_{\nu]} \nonumber \\
\left[M_{\mu\nu},M_{\alpha \beta}   \right]&=& i \left( \eta_{\alpha [\nu}M_{\mu]\beta}
+ \eta_{\beta [\mu}M_{\nu] \alpha}  \right) ~.
\label{commutatorsPoincare}
\end{eqnarray}
The main element of novelty imposed by the noncommutativity is found in the description
of the action of symmetry generators on products of fields: in light of the
noncommutativity of $\theta$-Minkowski coordinates, governed by (\ref{canonical}),
one has a
Baker-Campbell-Hausdorff formula, for the product of the exponentials in the
basis, of the familiar
form $e^{i k^\mu x_\mu}e^{i q^\mu x_\mu}
= e^{-\frac{i}{2} \theta_{\mu\nu} k^\mu q^\nu} e^{i (k^\mu + q^\mu) x_\mu}$,
and this in turn implies that the action of symmetry generators on products of fields
cannot obey a standard Leibniz rule.
Such anomalies in the  Leibniz rule are the core feature for which Hopf algebras
are equipped, describing them in terms of the so-called ``co-products".
From (\ref{actionP}) and (\ref{actionM}) it follows that
\begin{equation}
\begin{split}
P_\mu \left( e^{i k x} e^{i q x} \right) & = \left( P_\mu e^{i k x} \right) e^{i q x} + e^{i k x} \left( P_\mu e^{i q x} \right) ~, \\
M_{\mu\nu} \left( e^{i k x} e^{i q x} \right) & = \left( M_{\mu\nu} e^{i k x} \right) e^{i q x} + e^{i k x} \left( M_{\mu\nu} e^{i q x} \right) \\ & - \tfrac{1}{2} \theta^{\alpha\beta} \left[ \eta_{\alpha[\mu} \left( P_{\nu]} e^{ikx} \right) \left( P_{\beta} e^{iqx} \right) + \left( P_{\alpha} e^{ikx} \right) \eta_{\beta[\mu} \left( P_{\nu]} e^{iqx} \right) \right] ~.
\label{DefLeibniz}
\end{split}
\end{equation}
which in Hopf-algebra jargon amounts to a primitive coproduct for translations,
but a non-primitive (non-cocommutative) coproduct in the Lorentz sector:
\begin{eqnarray}
\Delta P_\mu&=&P_\mu\otimes \mathbbm{1}
+  \mathbbm{1} \otimes P_\mu~, \nonumber \\
\Delta M_{\mu \nu}&=& M_{\mu \nu}\otimes \mathbbm{1}
+  \mathbbm{1} \otimes M_{\mu \nu} -\frac{1}{2}\theta^{\alpha\beta}\left( \eta_{\alpha [\mu}
P_{\nu]}\otimes P_\beta +P_\alpha\otimes \eta_{\beta [\mu}P_{\nu]}\right) ~,
\label{Coproducts}
\end{eqnarray}
Let us note in passing that it is of some mathematical
interest
that this $\theta$-Poincar\'e Hopf algebra, characterized by the coproducts (\ref{Coproducts}),
can be described as the result of ``twisting" of the classical Poincar\'e algebra
by the twist element~\cite{chaichan,wessfiore,balaIRUV,wessNCYM,thetanoether}
\begin{equation}
\mathcal F = e^{\frac{i}{2} \theta^{\mu\nu}P_\mu \otimes P_\nu }.
\label{twistelement}
\end{equation}

The hypothesis that this ``twisted" $\theta$-Poincar\'e Hopf algebra should describe the symmetries
of $\theta$-Minkowski finds preliminary support in the observation that
the $\theta$-Minkowski commutation relations (\ref{canonical}) are invariant under the
action of the generators of $\theta$-Poincar\'e
\begin{equation}
\begin{array}{c}
P_\mu \triangleright [x_\rho , x_\sigma ] = (P_\mu \triangleright  x_{[\rho}) x_{\sigma]} + x_{[\rho} (P_\mu \triangleright x_{\sigma]} ) = i \eta_{\mu[\rho} x_{\sigma]} + i \eta_{\mu[\sigma} x_{\rho]} = 0 \equiv P_\mu \triangleright ( i \theta_{\rho\sigma}) ~, \\\\
M_{\mu\nu} \triangleright [x_\rho , x_\sigma ] = (M_{\mu\nu}  \triangleright  x_{[\rho}) x_{\sigma]} + x_{[\rho} (M_{\mu\nu}  \triangleright x_{\sigma]} ) - \frac{1}{2}\theta^{\alpha\beta}\left[ \eta_{\alpha [\mu}(
P_{\nu]} \triangleright x_{[\rho}) (P_\beta \triangleright x_{\sigma]}) + (P_\alpha\triangleright x_{[\rho}) \eta_{\beta [\mu} (P_{\nu]} \triangleright x_{\sigma]})\right] =\vspace{6pt} \\
 i ([x^{\alpha}, x^{\beta}]  - i  \theta^{\alpha\beta} )\eta_{\alpha [\mu}
\eta_{\nu] [\rho}\eta_{\sigma] \beta} = 0 \equiv M_{\mu\nu}  \triangleright ( i \theta_{\rho\sigma})~,
\end{array}
\label{maininvariance}
\end{equation}
where we also made use of the coproducts (\ref{Coproducts}) and of the rules of action (\ref{actionP}) and (\ref{actionM}) applied on the coordinates.

We shall give more substance to the claim that the $\theta$-Poincar\'e Hopf algebra
 describes the symmetries
of $\theta$-Minkowski in the next sections. Before doing that we find appropriate to
revisit the {\it ad hoc}, but evidently efficacious, introduction
of ``infinitesimal noncommutative parameters" for the purpose of
obtaining conserved currents from the Hopf-algebra
symmetries~\cite{kappanoether,jurekNOETHER1,jurekNOETHER2,thetanoether,nopure}.
For the $\theta$-Poincar\'e case this prescription amounts to~\cite{thetanoether}
\begin{equation}
df(x)=i \left[\gamma^\alpha P_\alpha
+ \omega^{\mu\nu}M_{\mu\nu}\right]f(x) ~,
\label{tuttodf}
\end{equation}
with transformation parameters that
 act on the spacetime coordinates only by (associative) multiplication and with rules
 for products of transformation parameters and spacetime coordinates such that
\begin{eqnarray}
\left[x^\beta,\gamma^\alpha \right]&
=&-\tfrac{i}{2}\omega^{\mu\nu}(\theta_{[\mu}\,^\alpha \delta_{\nu]}\,^\beta
+ \theta^\beta\,_{[\mu} \delta_{\nu]}\,^\alpha) ~, \label{nopureLor} \\
\left[x^\beta,\omega^{\mu\nu}\right]&=& 0 ~.\nonumber
\end{eqnarray}
It is then easy to verify~\cite{thetanoether},
using (\ref{nopureLor}), that the $df$ given in (\ref{tuttodf}) is a legitimate
differential, which satisfies
 Leibniz rule
\begin{equation}
d(f(x)g(x))=(df(x))g(x)+f(x)(dg(x)) ~.
\label{leibniz}
\end{equation}

The details of how this allows a successful Noether analysis can be found
in Ref.~\cite{thetanoether}.
We shall here just recall that
for a theory of
free massless scalar fields
in $\theta$-Minkowski with equation of motion
\begin{equation}
\Box \phi(x) \equiv P_\mu P^\mu \phi(x) = 0
\label{equationofmotionFree}
\end{equation}
this type of Noether analysis produces~\cite{thetanoether}
the following Noether current for translation symmetry
\begin{equation}
T_{\mu\nu} = \eta_{\mu\nu} \mathcal{L} - \left( P_\mu \phi(x) \right) P_\nu \phi(x) - \left( P_\nu \phi(x) \right) P_\mu \phi(x) ~,
\label{tmunuZERO}
\end{equation}
and the following Noether current for Lorentz-sector symmetry
\begin{equation}
{K^\alpha}_{\mu\nu} = \tfrac{1}{2} \{ x_{[\mu} , {T^\alpha}_{\nu]} \} ~.
\label{kZERO}
\end{equation}

\section{Quantum Poincar\'e groups}\label{quantumgroup}
The description of transfomations of fields given in the previous section
has the merit of producing a working Noether analysis, but does not match
 any previously-known
formalization of quantum-group symmetry transformations.
The resulting Noether charges (obtained by
integration~\cite{kappanoether,jurekNOETHER1,jurekNOETHER2,thetanoether,nopure}
of the Noether currents) are indeed conserved,
so the procedure does work.
But the attribution of the conserved charges to the
Hopf-algebra symmetries
remains dubious because of the mediation of the ``foreign" transformation parameters.
In this section we review (and elaborate on) the description of field transformations
given genuinely in terms of quantum-group structures. This will prepare the ground for
the result we shall seek in the next section, which is a  Noether
analysis relying exclusively on a genuine quantum-group description of
field transformations.

We start, in the first subsection, by describing the classical Poincar\'e group
as a Hopf algebra, since that will render more intelligible the picture
of quantum-group transformations discussed in the following subsections.

\subsection{The classical Poincar\'e group as a Hopf algebra}

The Poincar\'e group $\G$ is
 of course defined by its associative product law, its identity element, and its inverse,
\begin{equation}
\begin{array}{c}
(a,\Lambda) \cdot (a',\Lambda' ) = (\Lambda \cdot a' + a,\Lambda \cdot \Lambda' ) \qquad \forall ~ (a,\Lambda)  \in \G ~,
\\\\
(a,\Lambda) \cdot (0,I ) = (0,I ) \cdot (a,\Lambda) = (a,\Lambda) ~,
\\\\
(a,\Lambda) \cdot (- \Lambda^T \cdot a ,\Lambda^T ) = (- \Lambda^T \cdot a,\Lambda^T ) \cdot (a,\Lambda) =  (0,I ) ~,
\end{array}
\end{equation}
where $\Lambda$ represents the Lorentz matrices, $a$ the translation vectors, and $I$ the identity matrix.
The algebra $\CG$ is the unital commutative algebra of functions over the Poincar\'e group, with pointwise product and trivial linear space structure,
\begin{equation}
\begin{array}{c}
(f \cdot g) (X) = f(X) g(X) \qquad  \forall ~ X \in \G ~,
\\\\
(\alpha f + \beta g) (X) = \alpha f(X) + \beta g(X) ~, \label{CommutativeFunctionAlgebra}
\end{array}
\end{equation}
and identity given by $\id (X) = 1$. $\CG$ can encode the group structure of $\G$ if it is generalized to a Hopf algebra~\cite{AschieriQuantumPoincareGroup}
(also see Ref.~\cite{majidbook}),
\begin{equation}
\begin{array}{c}
 \Delta(f) (X,X') = f( X \cdot X') ~,
\\\\
 S(f) (X) = f( X^{-1}) ~,
\\\\
 \epsilon(f) = f(\id) ~.
\end{array}
\end{equation}
In fact from the coproducts in $\CG$ one can reconstruct the product law of $\G$, and from the antipodes and identity one deduces the inverses and the identity element.

A basis for $\mathbb{C}(\G)$ is given by the functions ${\Lambda^\mu}_\nu$ and $a^\mu$, which give the matrix entries of the defining representation of $\G$ when calculated over a group element. Their Hopf algebra structure is
\begin{equation}
\begin{array}{c}
 \Delta({\Lambda^\mu}_\nu)= {\Lambda^\mu}_\rho \otimes {\Lambda^\rho}_\nu ~, \qquad \Delta(a^\mu) = {\Lambda^\mu}_\nu \otimes a^\nu + a^\mu \otimes \id ~,
\\\\
 S({\Lambda^\mu}_\nu) = {(\Lambda^{-1})^\mu}_\nu ~, \qquad S(a^\mu) = - {(\Lambda^{-1})^\mu}_\nu a^\nu ~,
\\\\
 \varepsilon ({\Lambda^\mu}_\nu) = {\delta^\mu}_\nu ~, \qquad \varepsilon(a^\mu) = 0 ~.
\end{array} \label{GroupAlgebra}
\end{equation}

The classical Poincar\'e group is the group of the isometries of
classical Minkowski spacetime $\mathcal{M}$.
We can represent the coordinate systems over  $\mathcal{M}$ with the commutative algebra $\CM$ of functions over the Minkowski space. We'll represent the coordinate functions (which, when calculated over a point give its four coordinates in a certain coordinate system) as $x^\mu$, and we'll use them as a basis for our Hopf algebra.

The Hopf algebra structure of $\CM$ is
\begin{equation}
\begin{array}{c}
\Delta (x^\mu) = x^\mu \otimes \id + \id \otimes x^\mu ~, \qquad S(x^\mu) = - x^\mu~, \qquad \varepsilon (x^\mu) = 0 ~; \label{FunctionsOverMinkowski}
\end{array}
\end{equation}

A classical Poincar\'e transformation of the classical Minkowski spacetime $\mathcal{M}$
is often described as a map from the space-time coordinates $x^\mu$ to a new set of coordinates $x'^\mu = a^\mu + x^\nu  {\Lambda^\mu}_\nu$, with the transformed coordinates $x'^\mu$ given by elements of the (defining representation of the) Poincar\'e group multiplied and/or summed with spacetime coordinates $x^\mu$. In order to give a more robust description, suitable for generalization to the ``quantum symmetries" that are here of our interest,
 one must pay attention to the fact that
these transformations involve elements of different algebras, so one must specify what
 is meant by ``products'' and ``sums'' involving elements of these different algebras.
The simplest way to allow for sums and products of two algebras is the \emph{direct product} of the two algebras $\CM \otimes \CG$ (or, equivalently, $\CG \otimes \CM$). Within this framework, it is clear that the Poincar\'e transformation map goes from $\CM$ to the direct product $\CM \otimes \CG $ ( or $\CG \otimes \CM$); so it is a \emph{right (left) coaction},
\begin{equation}
\Delta_R : \CM \rightarrow \CM \otimes \CG  ~, \qquad  \Delta_R [x^\mu] = \id \otimes a^\mu + x^\nu \otimes {\Lambda^\mu}_\nu ~. \label{RightCoaction}
\end{equation}
The coaction is an algebra homomorphism with respect to the product of $\CM$:
\begin{equation}
\Delta_R[x^\mu x^\nu] = \Delta_R[x^\mu] \Delta_R[x^\nu] ~,
\end{equation}
so that it can be extended over polynomials in $\CM$, and consequently over all functions of $\CM$.

\subsection{General structure of Quantum Poincar\'e groups}
We have shown  a description of classical Poincar\`e symmetries such that all the group properties are codified in the coalgebraic sector of the Hopf algebra $\CG$.
In that familiar context the Hopf algebra, $\CG$, is commutative, but in general (particularly for the application to the description of the spacetime symmetries of quantum spacetimes, which are here of interest) this commutativity need not be enforced. One can indeed deform the algebraic structure of $\CG$,
$$
\CG \rightarrow \CqG
$$
making it into a \emph{noncommutative Hopf algebra}, or \emph{quantum group}, without modifying $\G$.
Such a deformation does not allow anymore the description of $\CqG$ in terms of (\ref{CommutativeFunctionAlgebra}), because the product there is necessarily commutative, but instead we can start from the basis  ${\Lambda^\mu}_\nu$, $a^\mu$, with the coalgebraic properties (\ref{GroupAlgebra}) and deform its commutators
$[a^\mu , a^\nu]$, $[a^\mu , {\Lambda^\rho}_\sigma]$, $[{\Lambda^\mu}_\nu , {\Lambda^\rho}_\sigma]$.

The commutation rules of $\CqG$ depend on those of the spacetime coordinates, because the noncommutativity of $\CqG$ is needed to enforce the homogeneity of spacetime.
In fact, from the homomorphism property of the coaction (\ref{RightCoaction}) one can see that the noncommutativity of the algebra  $\CqM$ implies the noncommutativity of $\CqG$, and vice versa.

Here we are interested in noncommutative \emph{algebraically homogeneous}
 spacetimes. This means that the algebraic properties are the same for all points of space-time. 
This requires us to impose that the coordinates $x_\mu$ of every point satisfy the same algebra $\CqM$:
\begin{equation}
 [x_\mu , x_\nu] = i ~ \Theta_{\mu \nu}(x) ~,
 \label{GeneralNoncommSpacetime}
\end{equation}
which means that $\Theta_{\mu \nu}(x)$ is the same function of the coordinates in every  coordinate system.

Since we want to preserve the commutation rules of spacetime under Poincar\'e transformations, in the noncommutative case the transformed coordinates,
$$
x'^\mu = \id \otimes a^\mu + x^\nu \otimes {\Lambda^\mu}_\nu ~ \in ~ \CqM \otimes \CqG ~,
$$
should be required to satisfy the same commutation rules as (\ref{GeneralNoncommSpacetime}),
\begin{equation}
 [x'^\mu , x'^\nu] = i ~ \Theta^{\mu\nu}(x') ~, \label{GeneralTransformedCommRules}
\end{equation}
so the coaction should still be a homomorphism with respect to the noncommutative product of $\CqM$
\begin{equation}
\Delta_R [f(x) g(x)] = \Delta_R[f(x)]\Delta_R[g(x)] ~. \label{HomomorphismProperty}
\end{equation}
In the case of a noncommutative space, $\Theta_{\mu \nu} (x) \neq 0$,
one can use this requirement to derive the noncommutativity of the
product of the generators of $\CqG$; in fact, imposing (\ref{GeneralTransformedCommRules}) one finds
\begin{equation}
\Theta_{\mu \nu} (x') = [x'^\mu , x'^\nu] = \id \otimes [a^\mu , a^\nu]  +  x^\rho \otimes \left( [a^{\mu} , {\Lambda^{\nu}}_\rho] - [ a^{\nu} , {\Lambda^{\mu}}_\rho ] \right) + x^\sigma  x^\rho \otimes \left( {\Lambda^{\mu}}_\rho {\Lambda^{\nu}}_\sigma - {\Lambda^{\nu}}_\rho {\Lambda^{\mu}}_\sigma \right) ~.
\end{equation}

The coaction is specified explicitly only over a single coordinate, but from the homomorphism property (\ref{HomomorphismProperty}) it can be computed over every product of coordinates. For example one has
\begin{equation}
 \Delta_R[x^\mu x^\nu] = \Delta_R[x^\mu]\Delta_R[x^\nu] = \id \otimes a^\mu a^\nu  +  x^\rho \otimes \left( a^{\mu}  {\Lambda^{\nu}}_\rho + {\Lambda^{\mu}}_\rho a^{\nu} \right) + x^\sigma  x^\rho \otimes {\Lambda^{\mu}}_\rho {\Lambda^{\nu}}_\sigma ~.
\end{equation}

The right coaction of the group over the algebra $\CqM$ of functions over the noncommutatve spacetime (understood as the universal enveloping algebra to the coordinate algebra (\ref{GeneralNoncommSpacetime})), gives an element of the tensor product $\CqM \otimes \CqG$,
$$
\Delta_R [f(x)] = f^{(1)}(x) \otimes f'^{(2)} ~, \qquad f^{(1)}_j(x)  \in \CqM ~, ~ f'^{(2)}_j \in \CqG ~,
$$
where we use Sweedler's notation with summation understood.
The above coaction can be viewed as a combination of \emph{actions} (linear maps from $\CqM$ to $\CqM$) over $f(x)$,
\begin{equation}
 \Delta_R[f(x)] = \sum_n T_n[f(x)] \otimes g_n ~, \qquad g_n \in \CqG ~, \; T_n: \CqM \rightarrow \CqM ~, \label{CoazioneFunzioneGiusta}
\end{equation}
where the $T_n$ are linear operators acting over the function $f(x)$, and the group elements $g_n$ are polynomials in the basis elements ${\Lambda^\mu}_\nu$ and $a^\mu$. The linear operators $T_n$ constitute a Hopf algebra, $U_q[\g]$, which is dual to $\CqG$.
In fact, a coaction of a Hopf algebra $\mathcal H$ over another Hopf algebra $\mathcal A$ induces an action of the dual Hopf algebra $\mathcal H^*$ over $\mathcal A$ (\cite{majidAlgebraic2,Zumino}):
$$
 h \triangleright a = a^{(1)} ~ (h , a'^{(2)}) ~, \qquad \forall ~ h \in \mathcal{H^*} ~, ~ a \in \mathcal{A}~,
$$

As suggested by our description of the commutative case, one should look for the algebra $U_q(\g)$ dual to the group $\CqG$ by developing the coaction over $\CqM$ in powers of the parameters (or group coordinates) $a_\mu$ and ${\omega^\mu}_\nu = {(\log \Lambda )^\mu}_\nu$. Considering the first order
\begin{equation}
 \Delta_R[f(x)] = f(x) \otimes \id + i P^\mu \triangleright [f(x)] \otimes a_\mu + \frac{i}{2}  M^{\mu\nu} \triangleright [f(x)] \otimes \omega_{\mu\nu}  + \mathcal{O} (a^2,\omega^2,a \omega)~,
\end{equation}
one infers that the operators $P^\mu$ and $M^{\mu\nu}$ should be our deformed-algebra generators, expected to be the dual Hopf algebra to $\CqG$, called $U_q(\g)$, with:
\begin{enumerate}
\item an action over $\CqM$ which is dual to the coaction of the 1-parameters subgroups of $\CqG$;
\item a coproduct that can be found by requiring compatibility with the product of $\CqM$, by acting on product of functions;
\item commutators that can be found by compatibility with the composition law of $\CqG$ (which is encoded into its coproducts);
\item antipodes and counits which are dual to those of $\CqG$.
\end{enumerate}
But it should be noticed that this procedure is affected essentially by an ordering ambiguity. Since the group parameters in general do not commute, it is not clear in what sense one can cut off the series expansion to first order in \emph{all} parameters: this is a meaningful concept in the commutative case, or in the case in which one has only one parameter, but not in the case in which one has products of noncommuting parameters. For example, in the case $[a_\mu , a_\nu ] =  i \alpha_{\mu\nu}$ with constant $\alpha_{\mu\nu}$,  one could naively describe the monomial $a_1 a_2$ as a second-order expression, but this does not take into account the fact that
$a_1 a_2 ~ = ~ a_2 a_1 + i \alpha_{12}$.
To give a sensible notion of ``first order'' it appears that one should necessarily adopt
 an \emph{ordering prescription}, for example:
$: a_2 a_1 : ~ = ~ \frac{1}{2} (a_1 a_2 + a_2 a_1)$, $: a^\rho \omega^{\mu\nu} : ~ = ~ \omega^{\mu\nu} a^\rho$,
so that one can write the coaction as an ordered power series:
\begin{equation}
\begin{array}{l}
\Delta_R \left[ f(x) \right] =
\sum_{n,m} \frac{(i)^{n+m}}{2^m n!m!} \bar M^{\rho_1 \sigma_1} \dots \bar M^{\rho_m \sigma_m}  \bar P^{\mu_1} \dots  \bar P^{\mu_n} \triangleright f(x) \otimes : a_{\mu_1} \dots a_{\mu_n} \omega_{\rho_1 \sigma_1} \dots \omega_{\rho_m \sigma_m}: ~,
\end{array}
\end{equation}
where, now, the algebra generators $\bar M^{\mu\nu}$ and $\bar P^\mu_w$ are ordering-choice-dependent.
Any change of ordering corresponds to a nonlinear change of algebra generators, which is perfectly legitimate in the universal enveloping algebra framework.

We can then find the coproducts for every basis of the algebra $U_q(\g)$ linked with an ordering choice for the coaction, by considering the right coaction over a product of functions and ask the homomorphism property,
\begin{eqnarray*}
&:e^{\frac{i}{2} \bar M^{\mu\nu} \otimes \omega_{\mu\nu}} e^{i \bar P^\rho \otimes a_\rho}: [f(x)g(x)\otimes \id] = &\\
&:e^{\frac{i}{2} \bar M^{\mu\nu} \otimes \omega_{\mu\nu}} e^{i \bar P^\rho \otimes a_\rho}:[f(x)\otimes \id] :e^{\frac{i}{2} \bar M^{\mu\nu} \otimes \omega_{\mu\nu}} e^{i \bar P^\rho \otimes a_\rho}: [g(x)\otimes \id] ~.&
\end{eqnarray*}
The role of the coproduct in the coaction over a product of functions can be made manifest in this way:
$$
\Delta _R[f(x)g(x)]=\mu _{12}\left\{:e^{\frac{i}{2} \Delta(\bar M^{\mu\nu}) \otimes \omega_{\mu\nu}} e^{i \Delta(\bar P^\rho) \otimes a_\rho}:[f(x)\otimes g(x)\otimes \id]\right\}
$$
where $ \mu _{12}: \CqM \otimes \CqM \otimes \CqG\longrightarrow \CqM \otimes \CqG$ is the representation in the triple tensor product of the product of $\CqM$: $\mu _{12}\left\{f(x)\otimes g(x)\otimes \alpha \right\} = f(x)\cdot g(x)\otimes \alpha  $, $\forall  ~ \alpha \, \in \, \CqG$; $\forall ~ f,g \, \in \, \CqM$. $\Delta$ is the coproduct of  $U_q(\g)$.

Imposing that the last expression is equal to $\Delta _R[f(x)]\Delta _R[g(x)]$ for all $f(x)$, $g(x)$, we can compute the coproducts,
\begin{equation}
:e^{\frac{i}{2} \Delta(\bar M^{\mu\nu}) \otimes \omega_{\mu\nu}} e^{i \Delta(\bar P^\rho) \otimes a_\rho}: \, = \, :e^{\frac{i}{2} \bar M^{\mu\nu} \otimes \id \otimes \omega_{\mu\nu}} e^{i \bar P^\rho \otimes \id  \otimes a_\rho}::e^{\frac{i}{2} \id \otimes  \bar M^{\mu\nu} \otimes \omega_{\mu\nu}} e^{i \id \otimes \bar P^\rho \otimes a_\rho}: \, . \label{CoproductsFromCoaction}
\end{equation}
From the equation above it is clear that if all the $a^\mu$ and the $\omega^{\mu\nu}$ commute, the coproducts will be primitive. In fact in that case, since $\id \otimes T$ and $T' \otimes \id$ commute for all $T$ and $T'$ in $U_q(\g)$, we get:
\begin{equation}
 :e^{\frac{i}{2} \Delta(\bar M^{\mu\nu}) \otimes \omega_{\mu\nu}} e^{i \Delta(\bar P^\rho) \otimes a_\rho}: \,= \, :e^{\frac{i}{2} (\bar M^{\mu\nu} \otimes \id + \id \otimes  \bar M^{\mu\nu} ) \otimes \omega_{\mu\nu}} e^{i ( \bar P^\rho \otimes \id +  \id \otimes \bar P^\rho ) \otimes a_\rho}: ~. \label{EqCoprdotto}
\end{equation}

One can also enforce that the coaction be a homomorphism also with respect to the composition law of the Poincar\'e group:
\begin{eqnarray*}
& :e^{\frac{i}{2} \bar M^{\mu\nu}  \otimes \omega_{\mu\nu}  } e^{i \bar P^\rho \otimes  a_\rho  } : [ \id \otimes ( a , \Lambda ) \cdot ( a' , \Lambda' )] = & \\
& : e^{\frac{i}{2} \bar M^{\mu\nu}  \otimes \omega_{\mu\nu}  } e^{i \bar P^\rho \otimes a_\rho } : [ \id  \otimes ( a , \Lambda )] : e^{\frac{i}{2} \bar M^{\mu\nu}  \otimes \omega_{\mu\nu}   } e^{i \bar P^\rho \otimes a_\rho } :[\id  \otimes (a' , \Lambda' ) ]&
\end{eqnarray*}
where the tensor product is understood as $U_q(g) \otimes \G $, since the group-elements $a_\rho$ and ${{\Lambda}^\mu}_\nu$ act as functions over $\G$.

We know that the coproduct of $\CqG$ reproduces the product in $\G$, so we can write
\begin{eqnarray*}
&&:e^{\frac{i}{2} \bar M^{\mu\nu}  \otimes \omega_{\mu\nu}  } e^{i \bar P^\rho \otimes  a_\rho  } : [ \id \otimes ( a , \Lambda ) \cdot ( a' , \Lambda' ) ] =
\\
&& \mu_{23} \left\{ :e^{\frac{i}{2} \bar M^{\mu\nu}  \otimes \Delta (\omega_{\mu\nu})  } e^{i \bar P^\rho \otimes  \Delta(a_\rho)  } : [ \id \otimes ( a , \Lambda ) \otimes ( a' , \Lambda' ) ] \right\}
\end{eqnarray*}
where $ \mu _{23}: U_q[\g] \otimes \G \otimes \G  \longrightarrow U_q(\g) \otimes \G$ is the representation in the triple tensor product of the product of $\G$: $\mu _{23}\left\{ T \otimes ( a , \Lambda ) \otimes ( a' , \Lambda' )  \right\} = T \otimes ( a , \Lambda ) \cdot ( a' , \Lambda' )$, $\forall ~  T  \, \in \, U_q[\g]$; $\forall  ~ ( a , \Lambda ) , ( a' , \Lambda' ) \, \in \, \G$.

This leads us to the following relationship
\begin{equation}
:e^{\frac{i}{2} \bar M^{\mu\nu}  \otimes \Delta(\omega_{\mu\nu} ) } e^{i \bar P^\rho \otimes \Delta( a_\rho ) } : ~ = ~ : e^{\frac{i}{2} \bar M^{\mu\nu}  \otimes \id \otimes \omega_{\mu\nu}   } e^{i \bar P^\rho \otimes \id \otimes a_\rho } : \,  : e^{\frac{i}{2} \bar M^{\mu\nu}  \otimes \omega_{\mu\nu}  \otimes \id } e^{i \bar P^\rho \otimes a_\rho \otimes \id } : \label{CometrovareCommutatori}
\end{equation}
where
$$
\Delta(a^\mu) = {\Lambda^\mu}_\nu \otimes a^\nu + a^\mu \otimes \id ~, \qquad \Delta(\omega^{\mu\nu} ) = \omega^{\mu \nu} \otimes \id + \id \otimes \omega^{\mu \nu} ~.
$$

For example, in the commutative case eq.(\ref{CometrovareCommutatori}) reduces to
\begin{equation}
\begin{array}{c}
e^{\frac{i}{2} M^{\mu\nu}  \otimes (\omega_{\mu \nu} \otimes \id + \id \otimes \omega_{\mu \nu} ) } e^{i P^\rho \otimes ( {\Lambda^\rho}_\sigma \otimes a^\sigma + a^\rho \otimes \id ) } = \\
e^{\frac{i}{2} M^{\mu\nu}  \otimes \id \otimes \omega_{\mu\nu}   } e^{i  P^\rho \otimes \id \otimes a_\rho } e^{\frac{i}{2} M^{\mu\nu}  \otimes \omega_{\mu\nu}  \otimes \id } e^{i P^\rho \otimes a_\rho \otimes \id }  ~,
\end{array}
\end{equation}
which is satisfied by the classical Poincar\'e algebra.

\subsection{Twisted Poincar\'e quantum group of symmetries of canonical spacetimes}
Let us now return to the ``canonical noncommutative spacetimes"
(or ``$\theta$-Minkowski"), characterized by
the canonical algebra $\Mtheta$:
\begin{equation}
 [x_\mu , x_\nu] = i \theta_{\mu \nu} ~.
\end{equation}
We shall describe the symmetries of these spacetimes
in terms of the twisted Poincar\'e quantum group~\cite{chaichan,KosinskiThetaPoincarGroup,oecklTHETA},
or ``$\theta$-Poincar\'e group", $\mathbb{C}_\theta[\G]$:
\begin{equation}
\begin{array}{c}
 \lbr a^{\mu },a^{\nu } \rbr = i \theta ^{\rho \sigma }\left({\delta^\mu}_\rho {\delta^\nu}_\sigma - {\Lambda^\mu}_\rho {\Lambda^\nu} _\sigma \right) ~, \qquad
\lbr {\Lambda^\mu}_\nu , {\Lambda^\rho} _\sigma \rbr = 0 ~, \qquad
\lbr {\Lambda^\mu}_\nu , a^\rho \rbr = 0
\\\\
\Delta({\Lambda^\mu}_\nu)= {\Lambda^\mu}_\rho \otimes {\Lambda^\rho}_\nu ~, \qquad \Delta(a^\mu) = {\Lambda^\mu}_\nu \otimes a^\nu + a^\mu \otimes \id ~,
\\\\
 S({\Lambda^\mu}_\nu) = {(\Lambda^{-1})^\mu}_\nu ~, \qquad S(a^\mu) = - {(\Lambda^{-1})^\mu}_\nu a^\nu ~,
\\\\
 \epsilon ({\Lambda^\mu}_\nu) = {\delta^\mu}_\nu ~, \qquad \epsilon(a^\mu) = 0 ~.
\end{array}
\end{equation}
It is important to notice that in $\theta$-Minkowski one cannot perform a pure
Lorentz transformation, a feature first emphasized in Ref.\cite{thetanoether}.
In fact,
from $\lbr a^{\mu },a^{\nu } \rbr = i \theta ^{\rho \sigma }\left({\delta^\mu}_\rho {\delta^\nu}_\sigma - {\Lambda^\mu}_\rho {\Lambda^\nu} _\sigma \right)$ it follows that
\begin{equation}
 \delta a^{\mu } \delta a^{\nu } \geq \left| \theta ^{\rho \sigma} \left\langle {\delta^\mu}_\rho {\delta^\nu}_\sigma - {\Lambda^\mu}_\rho {\Lambda^\nu} _\sigma \right\rangle \right| ~,
\end{equation}
and therefore a pure Lorentz transformation, which must be a transformation
with $\left\langle a_\mu \right\rangle = 0$, $\delta a^\mu = 0 $,
must necessarily be
trivial: $\left\langle {\Lambda^\mu}_\nu \right\rangle = {\delta^\mu}_\nu$.

For our purposes, most notably in the next section devoted to the Noether analysis,
it is sometimes convenient to consider the
expansion of the Lorentz matrix to first order:
\begin{equation}
 {\Lambda^\mu}_\nu \simeq \id + {\omega^\mu}_\nu + \mathcal{O} (\omega^2) ~,
\end{equation}
where $\omega_{\mu \nu} = - \omega_{\nu \mu}$.
This expansion is univocally defined, due to the commutativity of Lorentz group elements. And we shall occasionally make use of
\begin{equation}
 \lbr a^\mu , a^\nu \rbr = i \left(  \theta^{\nu \rho} {\omega^\mu}_\rho - \theta^{\mu \rho} {\omega^\nu}_\rho \right) \label{nopurethetaCONomega}
\end{equation}
which follows from $\lbr a^{\mu },a^{\nu } \rbr = i \theta ^{\rho \sigma }\left({\delta^\mu}_\rho {\delta^\nu}_\sigma - {\Lambda^\mu}_\rho {\Lambda^\nu} _\sigma \right)$.

%

It is straightforward to check that if the $U_\theta(\g)$ generators satisfy classical Poincar\'e commutation rules, the equation
\begin{equation}
 e^{\frac{i}{2}  M^{\mu\nu}  \otimes \Delta(\omega_{\mu\nu} ) } e^{i P^\rho \otimes \Delta( a_\rho ) }   =    e^{\frac{i}{2} M^{\mu\nu}  \otimes \id \otimes \omega_{\mu\nu}   } e^{i P^\rho \otimes \id \otimes a_\rho }  e^{\frac{i}{2} M^{\mu\nu}  \otimes \omega_{\mu\nu}  \otimes \id } e^{i   P^\rho \otimes a_\rho \otimes \id }  ~,
\end{equation}
is satisfied. In fact it is sufficient to commute the two exponentials in the middle of the \emph{rhs}
\begin{equation}
e^{i   P_\rho \otimes \id \otimes a^\rho }  e^{\frac{i}{2} M^{\mu\nu}  \otimes \omega_{\mu\nu}  \otimes \id } =  e^{\frac{i}{2} M^{\mu\nu}  \otimes \omega_{\mu\nu}  \otimes \id } e^{i   P_\rho \otimes {\Lambda^\rho}_\nu \otimes a^\nu }
\end{equation}
which is the classical formula for the adjoint action of a Lorentz transformation over a translation, and to observe that if $[P_\mu , P_\nu]=0$ then, since $a^\mu$ and ${\Lambda^\rho}_\sigma$ commute, we have
\begin{equation}
 e^{i   P_\rho \otimes {\Lambda^\rho}_\nu \otimes a^\nu }  e^{i   P_\rho \otimes a^\rho \otimes \id } = e^{i   P^\rho \otimes \Delta( a_\rho)} ~.
\end{equation}

Pure (both spatial and temporal) translations are allowed, so we can write:
\begin{equation}
\Delta_R[f(x)] = e^{i P_\mu \otimes a^\mu} \triangleright f(x) \otimes \id ~,
\end{equation}
assuming that ${\Lambda^\mu}_\nu = {\delta^\mu}_\nu$, which renders the $a^\mu$s commutative. This means that the coproducts of the $P_\mu$ are primitive:
\begin{equation}
 \Delta(P_\mu) = P_\mu \otimes \id + \id \otimes P_\mu ~.
\end{equation}

The nontrivial features are in the Lorenz sector:  imposing the homomorphism property to be satisfied by a complete transformation
we can compute the coproducts,
\begin{eqnarray}
&& e^{\frac{i}{2} \Delta(M_{\rho\sigma}) \otimes \omega^{\rho \sigma}} e^{i \Delta(P_\mu) \otimes a^\mu}[f(x)\otimes g(x)\otimes \id] = \nonumber
\\
&&e^{\frac{i}{2} M_{\rho\sigma} \otimes \id \otimes \omega^{\rho \sigma}} e^{i P_\mu \otimes \id \otimes a^\mu} e^{\frac{i}{2} \id \otimes  M_{\alpha \beta} \otimes \omega^{\alpha \beta}}  e^{i \id \otimes P_\nu \otimes a^\nu} [f(x)\otimes g(x)\otimes \id].
\end{eqnarray}
From this, using
$$
e^{i   P_\mu\otimes \id \otimes a^\mu}  e^{\frac{i}{2} \id\otimes M_{\alpha\beta} \otimes \omega^{\alpha \beta}}  =    e^{\frac{i}{2} \id\otimes M_{\alpha\beta} \otimes \omega^{\alpha \beta}}   e^{i   P_\mu\otimes \id \otimes a^\mu}  \,\, ,
$$
 which follows form the fact that $a^\mu$ and $\omega^{\rho \sigma}$ commute,
we obtain (using also the commutativity between the $\omega^{\rho\sigma}$)
\begin{equation}
e^{\frac{i}{2} \Delta(M_{\rho\sigma}) \otimes \omega^{\rho \sigma}} e^{i \Delta(P_\mu) \otimes a^\mu} =  e^{\frac{i}{2} (M_{\rho\sigma} \otimes \id + \id \otimes M_{\rho\sigma}) \otimes \omega^{\rho \sigma}} e^{i P_\mu \otimes \id \otimes a^\mu} e^{i \id \otimes P_\nu \otimes a^\nu}
\end{equation}
Then using the Baker-Campbell-Hausdorff formula one finds that
\begin{eqnarray*}
&& e^{i \Delta(P_\mu) \otimes a^\mu}  \equiv e^{i P_\mu \otimes \id \otimes a^\mu + i \id \otimes P_\mu \otimes a^\mu} = e^{i P_\mu \otimes \id \otimes a^\mu} e^{i \id \otimes P_\mu \otimes a^\mu} e^{\frac{1}{2} P_\mu \otimes P_\nu \otimes[a^\mu , a^\nu] }  \\
&&= e^{i P_\mu \otimes \id \otimes a^\mu} e^{i \id \otimes P_\mu \otimes a^\mu} e^{\frac{i}{2} \theta ^{\rho \sigma }  P_\mu \otimes P_\nu \otimes \left({\delta^\mu}_\rho {\delta^\nu}_\sigma - {\Lambda^\mu}_\rho {\Lambda^\nu} _\sigma \right)}\\
&&=e^{\frac{i}{2} \theta ^{\rho \sigma }  P_\mu \otimes P_\nu \otimes \left({\delta^\mu}_\rho {\delta^\nu}_\sigma - {\Lambda^\mu}_\rho {\Lambda^\nu} _\sigma \right)}
 e^{i P_\mu \otimes \id \otimes a^\mu} e^{i \id \otimes P_\mu \otimes a^\mu}
  \end{eqnarray*}
 where the last equality follows from the commutativity of the $P_{\mu}$'s.
So we finally arrive at the relation
\begin{equation}
e^{\frac{i}{2} \Delta(M_{\rho\sigma}) \otimes \omega^{\rho\sigma}}= e^{\frac{i}{2} \left( M_{\rho\sigma} \otimes \id  +  \id \otimes M_{\rho\sigma}\right ) \otimes \omega^{\rho\sigma} } e^{-\frac{i}{2}  \theta ^{\rho \sigma }  P_\mu \otimes P_\nu \otimes \left( \delta^{\mu}_\rho {\delta^\nu}_\sigma - {\Lambda^\mu}_\rho {\Lambda^\nu} _\sigma \right) } ~,
\end{equation}
that can be rewritten in the form
\begin{equation}
e^{\frac{i}{2} \Delta(M_{\rho\sigma}) \otimes \omega^{\rho\sigma}}e^{\frac{i}{2}\theta^{\mu\nu}P_{\mu}\otimes P_{\nu}\otimes \id}= e^{\frac{i}{2} \left( M_{\rho\sigma} \otimes \id  +  \id \otimes M_{\rho\sigma}\right ) \otimes \omega^{\rho\sigma} } e^{\frac{i}{2}  \theta ^{\rho \sigma }  P_\mu \otimes P_\nu \otimes {\Lambda^\mu}_\rho {\Lambda^\nu} _\sigma } ~.
\end{equation}
The last step consists in observing that
\begin{equation}
e^{\frac{i}{2} \id\otimes M_{\rho\sigma}\otimes\omega^{\rho\sigma}}e^{\frac{i}{2}\theta^{\rho\sigma} P_{\mu}\otimes P_{\nu}\otimes  {\Lambda^\mu}_\rho {\Lambda^\nu} _\sigma }e^{- \frac{i}{2} \id\otimes M_{\rho\sigma}\otimes\omega^{\rho\sigma}}=e^{\frac{i}{2}\theta^{\rho\sigma} P_{\mu}\otimes P_{\alpha}\otimes {{(\Lambda^{-1})}^\alpha}_\nu  {\Lambda^\mu}_\rho {\Lambda^\nu} _\sigma }
\end{equation}
so that we get an expression in terms of the twist element $\mathcal{F}$ in (\ref{twistelement})
\begin{equation}
e^{\frac{i}{2} \Delta(M_{\rho\sigma}) \otimes \omega^{\rho\sigma}}=  e^{\frac{i}{2}  \theta ^{\mu \nu }  P_\mu \otimes P_\nu \otimes \id} e^{\frac{i}{2} \left( M_{\rho\sigma} \otimes \id  +  \id \otimes M_{\rho\sigma}\right ) \otimes \omega^{\rho\sigma} } e^{-\frac{i}{2}\theta^{\mu\nu}P_{\mu}\otimes P_{\nu}\otimes \id}~.
\end{equation}
This, exploiting the ``twisting'' relation \cite{chaichan} $\Delta {M_{\mu\nu}}=\mathcal{F}(M_{\mu\nu}\otimes \id+\id \otimes M_{\mu\nu})\mathcal{F}^{-1}$, leads to
\begin{equation}
\Delta(M_{\rho\sigma}) = M_{\rho\sigma} \otimes \id +  \id \otimes M_{\rho\sigma} - \frac{1}{2} \theta^{\mu\nu} \left( \eta_{\mu [\rho} P_{\sigma]} \otimes P_\nu + P_\mu \otimes \eta_{\nu [\rho} P_{\sigma]} \right).
\end{equation}

\section{Noether analysis}\label{noether}
Equipped with the preparatory discussion reported in the previous section,
we can now turn to the main objective of this study, which concerns a Noether
analysis of the Hopf-algebra symmetries of $\theta$-Minkowski.
We shall describe as \emph{variation of the lagrangian} the following element of the tensor product $\CqM \otimes \CqG$
\begin{equation}
 \delta \Lag (x) = \Lag(\Delta_R[x]) - \Delta_R[\Lag(x)] ~ \in ~ \CqM \otimes \CqG~.
\end{equation}
If $\delta\Lag = 0$ then the lagrangian is a scalar field.

The variation of the lagrangian, as every element of the tensor product $\CqM \otimes \CqG$, can be expanded in powers of the transformation parameters $a^\mu$ and $\omega^{\mu_\nu}$, upon adopting an ordering prescription,
\begin{equation}
\delta\Lag = \mathfrak{P}_\mu  \otimes a^\mu + \mathfrak{M}_{\mu\nu} \otimes \omega^{\mu\nu} + \dots
\end{equation}
For an invariant lagrangian this must vanish, independently on the state of the transformation parameters $a^\mu$ and $\omega^{\mu\nu}$.
This must hold for every possible ordering choice for the transformation parameters, and therefore the invariance of the lagrangian implies infinitely many equations of the kind 
of $\mathfrak{P}_\mu = 0$ and $\mathfrak{M}_{\mu\nu}=0$.
However one must expect that the different laws found by changing ordering
 prescription are not independent from each other.

Of course, in the commutative case through the Noether analysis
one finds quantities $\mathfrak{P}_\mu $ and $\mathfrak{M}_{\mu\nu}$
which are both 4-divergences:
\begin{equation}
\mathfrak{P}_\mu = \partial_\nu T^{\mu\nu} ~, \qquad \mathfrak{M}_{\mu\nu} = \partial_\rho K^\rho_{\mu\nu} ~,
\end{equation}
so that from the fact that they vanish one obtains local conservation laws.

We shall now establish how these properties found in the commutative case
generalize to the case of noncommutative theories. We focus for simplicity
on the case of free scalar fields in $\theta$-Minkowski, for which a standard choice of
Lagrangian density is~\cite{thetanoether,wessfiore}
$$
 \Lag(x) = \frac{1}{2} \left\{ P_\mu \Phi(x) P^\mu \Phi(x) - m^2 \Phi^2(x) \right\} ~.
$$
This choice is motivated by the fact that the commutators among generators
of the $\theta$-Poincar\'e Hopf algebra are undeformed (the $\theta$-Poincar\'e
deformation is all in the coalgebra sector).

Then describing the variation of the lagrangian
as $\delta \Lag \simeq \mathfrak{P}_\mu  \otimes a^\mu + \mathfrak{M}_{\mu\nu} \otimes \omega^{\mu\nu}$
one easily finds a result for the translation sector of the $\theta$-Poincar\'e symmetries,
which is
$$ 0 = \mathfrak{P}^\mu  =  P_\nu T^{\mu\nu} $$
with
\begin{equation*}
 T^{\mu\nu} (x) = \frac{1}{2} \left\{   P^\nu  \Phi(x)  P^\mu \Phi(x) + P^\mu  \Phi(x)  P^\nu \Phi(x) \right\} - \eta^{\mu\nu} \Lag(x) ~.
\end{equation*}
For the Lorentz sector obtaining a fully intelligible result is slightly more tedious.
We first notice that
\begin{eqnarray}
0=\mathfrak{M}_{\rho\sigma} &\simeq&
 \frac{i}{4} \left\{ P_\mu  M_{\rho\sigma}  \Phi(x)  P^\mu \Phi(x) + P_\mu  \Phi(x)  P^\mu M_{\rho\sigma} \Phi(x) \right\} +\nonumber\\
&& \frac{i}{4} \left\{ \frac{1}{2}
 \Upsilon^{\alpha\beta}_{\rho\sigma} P_\mu  P_\alpha  \Phi(x)  P^\mu P_\beta \Phi(x) - M_{\rho\sigma}\left(P_\mu \Phi(x) P^\mu \Phi(x)\right) \right\} ~,
 \label{mainTHETA}
\end{eqnarray}
where we used the compact notation
$$
\Upsilon^{\alpha\beta}_{\rho\sigma}  =    {\theta^\alpha}_{[\sigma} {\delta^\beta}_{\rho]} - {\theta^\beta}_{[\sigma} {\delta^\alpha}_{\rho]} ~,
$$
so that in particular
$$
\Upsilon^{\alpha\beta}_{\rho\sigma} P_\alpha \otimes P_\beta  =   - \theta^{\mu\nu} \left(  \eta_{\mu[\rho} P_{\sigma]} \otimes P_\nu  +  P_\mu \otimes \eta_{\nu[\rho} P_{\sigma]}  \right) ~,
$$
and we recall that the coproduct of $ M_{\rho\sigma}$ takes the form
$$
\Delta(M_{\rho\sigma}) = M_{\rho\sigma} \otimes \id + \id \otimes M_{\rho\sigma} + \frac{1}{2} \Upsilon^{\alpha\beta}_{\rho\sigma} P_\alpha \otimes P_\beta ~.
$$

Our main objective is already achieved by Eq.~(\ref{mainTHETA}), which reproduces
the key point of the Noether analysis of scalar fields in $\theta$-Minkowski that
was previously obtained in Ref.~\cite{thetanoether}. Here this result has been
derived from an intelligible characterization of the quantum-group symmetry transformation
governed by $\theta$-Poincar\'e, while in  Ref.~\cite{thetanoether} one could
only obtain this result by the {\it ad hoc} introduction of the peculiar
transformation parameters on which we commented already in Section II.

With straightforward but tedious manipulations one can rearrange Eq.~(\ref{mainTHETA})
in the form
\begin{eqnarray*}
0=\mathfrak{M}_{\rho\sigma} &\simeq& \frac{i}{8} P_\mu \left\{ x_{[\sigma}  \left( P_{\rho]}  \Phi(x)  P^\mu \Phi(x) +  P^\mu  \Phi(x)   P_{\rho]} \Phi(x) - 2 {\delta^\mu}_{[\rho} \Lag(x) \right) \right\}+ \\
&& \frac{i}{8} P_\mu \left\{ \left( P_{[\rho}  \Phi(x)  P^\mu \Phi(x) +  P^\mu  \Phi(x)   P_{[\rho} \Phi(x)  - 2 {\delta^\mu}_{[\rho} \Lag(x)\right) x_{\sigma]} \right\}
~,
\end{eqnarray*}
which in turn can be usefully expressed in terms
of the energy-momentum tensor:
$$
0=\mathfrak{M}_{\rho\sigma} = \frac{i}{2} P_\mu \left(x_{[\rho} {T^\mu}_{\sigma]}(x) + {T^\mu}_{[\sigma}(x) x_{\rho]} \right)~.
$$
As expected we therefore find that the Noether analysis leads us to the introduction
of a conserved angular-momentum tensor $K^\mu_{\rho \sigma}$,
such that $\mathfrak{M}_{\rho\sigma} =  P_\mu K^\mu_{\rho \sigma} $,
and
$$
K^\mu_{\rho \sigma} (x) = \frac{1}{2} \left(x_{[\rho} {T^\mu}_{\sigma]}(x) + {T^\mu}_{[\sigma}(x) x_{\rho]} \right) ~.
$$

\section{Charges and stationary fields with space/time noncommutativity}\label{timeindep}
With the analysis reported so far, which was our main objective
for this manuscript,
we established more firmly and intelligibly
the attribution of conserved currents
and charges to the Hopf-algebra spacetime symmetries
of noncommutative field theories.
We feel that, to some extent,
having achieved this higher level of confidence in the possibility
of Noether analysis of these novel descriptions of spacetime symmetries,
we must now consider more urgent the investigation of other
aspects of these Noether-analysis results that are still not fully clarified.
The most urgent of these open issues concern the interpretation
of the conserved currents and charges produced by the Noether analysis.

It was already stressed in previous
works~\cite{kappanoether,jurekNOETHER1,jurekNOETHER2,thetanoether,nopure}
that at present the only tangible source of confidence in the criteria for current
conservation that are being adopted ($P_\mu J^\mu =0$ for $\theta$-Minkowski)
come from the fact that the associated conserved charges actually "work",
they are time independent when evaluated on solutions of the equation of motion.
So a crucial result for this whole research programme is the one reported
in detail
in Refs.~\cite{kappanoether,thetanoether,nopure},
 where the time independence of the conserved charges was verified.

 But in connection with the time independence of the conserved charges
 we feel that a subtle issue needs to be addressed. In order to introduce
 this issue let us quickly summarize
the analysis reported in Ref.\cite{thetanoether}
for the time independence of the energy-momentum charges
for solutions of the equation of motion in $\theta$-Minkowski.
These charges are of course obtained in terms of the energy-momentum tensor
in $\theta$-Minkowski spacetime (\ref{tmunuZERO})
through the formula: 
\begin{equation}
Q^P_\nu = \int d^3 x T_{0 \nu} ~,
\end{equation}
where the standard (and elementary) concept of spatial integration in such noncommutative
spacetimes (see, {\it e.g.}, Ref.~\cite{thetanoether}) is understood.

The charges $Q^P_\nu$ can be conveniently analyzed exploiting the
Fourier representation of
solutions of the equation of motion (\ref{equationofmotionFree}):
\begin{equation}
\Phi(x) = \int d^4k  ~\tilde \Phi(k) ~ \delta (k_\mu k^\mu) ~ e^{i k_\nu x^\nu} ~,
\end{equation}
where the form of the equation of motion is of course codified in $\delta (k_\mu k^\mu)$.
With this description of the solutions of the equation of motion and
the result (\ref{tmunuZERO})
for the energy-momentum tensor
in $\theta$-Minkowski spacetime one easily finds for the energy-momentum
charges $Q^P_\nu$ that
\begin{eqnarray}
Q^P_\nu &=&  \frac{1}{2} \int d^4k~ d^4q ~ \tilde \Phi(k) \tilde \Phi(q) \delta(k^2) \delta(q^2) \left(  {\delta^0}_\nu  k_\rho  q^\rho -  k^0  q_\nu - k_\nu  q^0  \right)  \int d^3 x ~ e^{i (k+q)_\alpha x^\alpha - \frac{1}{2} \theta^{\alpha\beta} k_\alpha q_\beta}  \nonumber \\
&=& \frac{1}{2}  \int d^4k~ d^4q ~ \tilde \Phi(k) \tilde \Phi(q) \delta(k^2) \delta(q^2) \left(  {\delta^0}_\nu  k_\rho  q^\rho -  k^0  q_\nu - k_\nu  q^0  \right) e^{i (k+q)_\alpha x^\alpha - \frac{1}{2} \theta^{\alpha\beta} k_\alpha q_\beta} \delta^{(3)}(\vec k + \vec q ) \nonumber
 \\
&=& \frac{1}{2}  \int d^4k~ d^4q~ \tilde \Phi(k) \tilde \Phi(q_0,-\vec k) \delta(k^2) \delta(q_0^2-|\vec k |^2) \left( \begin{array}{c}
k_0 q_0 - |\vec k|^2 \\
(q_0-k_0) k_j
\end{array}
 \right) e^{i (k+q)_0 x^0 - \frac{1}{2} \theta^{0j} (k_0-q_0) k_j } \delta^{(3)}(\vec k + \vec q ) ~, \label{indipendenza}\\
&=&  \int d^3k \frac{1}{4 |\vec k|}~ \tilde \Phi( |\vec k |,\vec k) \tilde \Phi(-|\vec k |,-\vec k) \delta(k_0 - |\vec k |) \left( \begin{array}{c}
- |\vec k| \\
 - k_j
\end{array}
 \right) e^{- \theta^{0j} |\vec k | k_j }  +
 \nonumber \\
&& \int d^3k  \frac{1}{4 |\vec k|}~ \tilde \Phi(-|\vec k |,\vec k) \tilde \Phi(|\vec k |,-\vec k) \delta(k_0 +|\vec k |) \delta(q_0 - |\vec k |) \left( \begin{array}{c}
 - |\vec k| \\
 k_j
\end{array}
 \right) e^{ \theta^{0j}  |\vec k | k_j }
  ~, \nonumber
\end{eqnarray}
which is explicitly time ($x_0$) independent.
Or is it not?
The steps of derivation we highlighted in (\ref{indipendenza}) show that
all sources of possible $x_0$ dependence cancel out, but is this sufficient
for establishing that a charge in a noncommutative spacetime is time independent.
The main source of our concerns comes from the case of "space/time
noncommutativity, {\it i.e.} when the time coordinate is itself noncommutative.
Think for example of the possibility of replacing $Q^P_\nu$
with $Q^P_\nu + i \theta_{01} + x_1 x_0 -x_0 x_1$:
of course the added term vanishes, $i \theta_{01} + x_1 x_0 -x_0 x_1 =0$,
but it also "depends on $x_0$" in the naive sense that it can be written
by formally introducing $x_0$.

The naivety of this example of$\,$\footnote{A similarly naive example is
the ``time dependence of $1$": $1=i(x_1 x_0 -x_0 x_1)/\theta_{01}$.} ``time dependent $0$", which we used to illustrate
our concerns, should not lead to underestimating the issue. At least in the physics
literature on spacetime noncommutativity concepts such
as "time-independent charge" and "stationary fields" have been used as if
time independence could be established "by inspection". We simply observe
that "time independence by inspection" could be misleading. On the other hand
what essentially we are going to argue is that one can rely
on "time independence by inspection" if all steps of the analysis have
been performed consistently with a given choice of ordering prescription
(this is of course what one typically does anyway, for independent reasons, in
physics applications of spacetime noncommutativity, and this is the reason
why no puzzling results on time independence were ever discussed).

We propose that time independence (as considered for example in the
characterization of conserved charges and stationary fields)
can be established if one writes the field of interest as an ordered
polynomial and finds that the time coordinate does not appear in the expression of the
field.

In order to explore the robustness of this definition it is useful to focus
on two examples of ordering convention and of corresponding Weyl maps,
and consider what they imply for our  example of "time dependent $0$".
With time-to-the-right ordering convention
\begin{equation}
: x_j x_0: \rightarrow x_j x_0 ~,~~~
:x_0 x_j: \rightarrow x_j x_0 + i \theta_{0j}~, \nonumber
\end{equation}
and
\begin{equation}
x_j x_0 = \Omega_r (x_0 x_j) ~,~~~ x_0 x_j  = \Omega_r (x_0 x_j + i \theta_{0j}) ~,\label{MappaRight}
\end{equation}
so that
$$:[i \theta_{01} + x_1 x_0 -x_0 x_1]: =0
= \Omega_r^{-1}(i \theta_{01} + x_1 x_0 -x_0 x_1)\,\, ,$$
where $\Omega_r^{-1}$ is the inverse of the Weyl map, taking form noncommutative
variables to commutative ones.

Similarly, for the Weyl ordering convention
\begin{equation}
: x_j x_0: \rightarrow \frac{1}{2} (x_j x_0 + x_0 x_j + i \theta_{j0} )~,~~~
:x_0 x_j: \rightarrow \frac{1}{2} ( x_j x_0 + x_0 x_j  + i \theta_{0j})~, \nonumber
\end{equation}
and
\begin{equation}
x_j x_0 = \Omega_w (x_0 x_j  + \frac{i}{2} \theta_{j0} ) ~,~~~ x_0 x_j  = \Omega_w (x_0 x_j  +\frac{i}{2} \theta_{0j}) ~,\label{MappaWeyl}
\end{equation}
so that
$$:[i \theta_{01} + x_1 x_0 -x_0 x_1]: =0
= \Omega_w^{-1}(i \theta_{01} + x_1 x_0 -x_0 x_1)\,\, .$$

As further evidence of robustness of our definition of time independence
we invite our readers to contemplate the equivalent description of
Weyl maps given in Section II, centered on the ordering convention for the
basis of exponentials used in the Fourier characterization of a field.
Evidently, from that perspective our definition of time independence
can be equivalently described as the statement that, working consistently
within one such ordered basis of exponentials, the Fourier transform ${\tilde f}(k)$
of a time-independent field $f(x)$ should be such that
$${\tilde f}(k) \propto \delta(k_0)\,\, .$$
And the fact that this is a sensible definition of time independence
is confirmed by the known fact (see, {\it e.g.}, Refs.\cite{gacalexluna,kappanoether})
that adopting different ordering conventions for the basis of exponentials
(different Weyl maps) leads in general to different Fourier transforms,
but the Fourier transforms obtained for a given field  according to
different ordering conventions always agree in $k_0 =0$.

It is also important to stress that according to
our characterization of time independence
one finds that $P_0 f(x) = 0 $ for a time-independent field $f(x)$
in $\theta$-Minkowski:
\begin{equation}
f(x) = \int d^3 k ~ \tilde g_b (\vec k) \Omega_b ( e^{i k \cdot \vec x} ) ~, ~ \Rightarrow ~ P_0 f(x) = i \int d^3 k ~ \tilde g_b (\vec k) \Omega_b ( \partial_0 e^{i k \cdot \vec x} ) = 0 ~.
\end{equation}
And notice that also the reverse it true: if $P_0 f(x) = 0$ then
\begin{equation}
P_0 \left[ \int d^4 k ~ \tilde f_b (k) \Omega_b ( e^{i k_\mu x^\mu} )\right] = 0  ~, ~ \Rightarrow ~  \tilde f_b (k) k_0  e^{i k_\mu x^\mu}  = 0~.
\end{equation}
where the equality on the right-hand side
 is a commutative functional equation, whose solution is easily found as
\begin{equation}
\tilde f_b (k)  =  \tilde g_b (\vec k) \delta(k_0) ~.
\end{equation}

We are here focusing on $\theta$-Minkowski, but it is easy to check (details can be found in
Ref.~\cite{flaviotesi}) that
also for the other most studied noncommutative spacetime,
the mentioned \kM ,
our definition of time independence is applicable and equally robust.

\section{Closing remarks}\label{closing}
We have here contributed to the fast growing maturity of the
hypothesis of a Hopf-algebra description of spacetime symmetries.
We feel that our main result, showing that conserved charges can be obtained directly
from Hopf-algebra symmetry transformations, might have implications
that go beyond the specific context of the Noether analysis.
In particular, by uncovering this robust behavior of finite
quantum-group symmetry transformations we hope to provide encouragement
for their more direct application also to the description of other aspects
of these theories in noncommutative spacetimes.

Among these possible applications
we feel that a special mention is deserved for the investigation
of the connection between spin and statistics
in noncommutative theories,
which may lead to valuable opportunities for phenomenology and
has been very actively debated in recent times
(see, {\it e.g.}, Refs.~\cite{balaSPINStat,michantoSPINstat,lukieSPINstat,youngSPINstat}).

Within the confines of Noether-analysis applications it is noteworthy
that so far these have concerned
exclusively theories of classical fields in the noncommutative spacetimes,
but valuable results could perhaps be obtained in analogous studies
of classical point particles. Particularly for the case of
classical point particles in $\kappa$-Minkowski spacetime it appears
reasonable to hope that such an approach might reconcile
the alternative scenarios which have so far emerged
from analyses based on more indirect
arguments~\cite{gacmaj,leeLIMITEDinertial,michJUREworldlines}.

Concerning a proper definition of stationary fields and conserved charges,
which we examined in detail here in Section~\ref{timeindep},
we feel that an interesting conceptual issue that one could contemplate is whether
the availability of such a sharp concept of time independence should be considered
necessary for applications in physics. We found that $\theta$-Minkowski
(and $\kappa$-Minkowski~\cite{flaviotesi}) do admit a sharp concept of time independence,
but this may well not be the case of other quantum spacetimes.


%

\end{document}